%% file: subleading.tex
\input myharvmac
\input amssym.def
\input amssym.tex
\noblackbox
\baselineskip=14.5pt

\def\comment#1{{}}

\newif\ifnref

\nreffalse

\input epsf

\def\figin{\epsfcheck\figin}\def\figins{\epsfcheck\figins}
\def\epsfcheck{\ifx\epsfbox\UnDeFiNeD
\message{(NO epsf.tex, FIGURES WILL BE IGNORED)}
\gdef\figin##1{\vskip2in}\gdef\figins##1{\hskip.5in}
\else\message{(FIGURES WILL BE INCLUDED)}%
\gdef\figin##1{##1}\gdef\figins##1{##1}\fi}
\def\DefWarn#1{}
\def\figinsert{\goodbreak\midinsert}  
\def\ifig#1#2#3{\DefWarn#1\xdef#1{Fig.~\the\figno}
\writedef{#1\leftbracket fig.\noexpand~\the\figno}%
\figinsert\figin{\centerline{#3}}\medskip\centerline{\vbox{\baselineskip12pt
\advance\hsize by -1truein\noindent\footnotefont\centerline{{\bf
Fig.~\the\figno}\ \sl #2}}}
\bigskip\endinsert\global\advance\figno by1}

\def\iifig#1#2#3#4{\DefWarn#1\xdef#1{Fig.~\the\figno}
\writedef{#1\leftbracket fig.\noexpand~\the\figno}%
\figinsert\figin{\centerline{#4}}\medskip\centerline{\vbox{\baselineskip12pt
\advance\hsize by -1truein\noindent\footnotefont\centerline{{\bf
Fig.~\the\figno}\ \ \sl #2}}}\smallskip\centerline{\vbox{\baselineskip12pt
\advance\hsize by -1truein\noindent\footnotefont\centerline{\ \ \ \sl #3}}}
\bigskip\endinsert\global\advance\figno by1}


\def\tilde{\widetilde}

\def\o {\over}
\def\fc#1#2{{#1 \o #2}}


\def\br{\hfill\break}

\def\Pc {{\cal P}}

\def\ceiling#1{\lceil#1\rceil}



\lref\KawaiXQ{
  H.~Kawai, D.C.~Lewellen and S.H.H.~Tye,
``A Relation Between Tree Amplitudes Of Closed And Open Strings,''
  Nucl.\ Phys.\  B {\bf 269}, 1 (1986).
}

\lref\StiebergerQJA{
  S.~Stieberger and T.R.~Taylor,
  ``Graviton Amplitudes from Collinear Limits of Gauge Amplitudes,''
Phys.\ Lett.\ B {\bf 744}, 160 (2015).
[arXiv:1502.00655 [hep-th]].
}

\lref\BjorkenDY{
  J.D.~Bjorken,
  ``Asymptotic Sum Rules at Infinite Momentum,''
Phys.\ Rev.\  {\bf 179}, 1547 (1969).
}

\lref\ParkeGB{
  S.J.~Parke and T.R.~Taylor,
  ``An Amplitude for $n$ Gluon Scattering,''
Phys.\ Rev.\ Lett.\  {\bf 56}, 2459 (1986).
}

\lref\AltarelliZS{
  G.~Altarelli and G.~Parisi,
  ``Asymptotic Freedom in Parton Language,''
Nucl.\ Phys.\ B {\bf 126}, 298 (1977).
}
\lref\StiebergerHBA{
  S.~Stieberger and T.R.~Taylor,
 ``Closed String Amplitudes as Single-Valued Open String Amplitudes,''
Nucl.\ Phys.\ B {\bf 881}, 269 (2014).
[arXiv:1401.1218 [hep-th]].
}

\lref\BernQJ{
  Z.~Bern, J.J.M.~Carrasco and H.~Johansson,
``New Relations for Gauge-Theory Amplitudes,''
Phys.\ Rev.\ D {\bf 78}, 085011 (2008).
[arXiv:0805.3993 [hep-ph]].
}

\lref\StiebergerHQ{
  S.~Stieberger,
``Open \& Closed vs. Pure Open String Disk Amplitudes,''
[arXiv:0907.2211 [hep-th]].
}

\lref\stnew{
  S.~Stieberger and T.R.~Taylor, in preparation.}

\lref\StiebergerCEA{
  S.~Stieberger and T.R.~Taylor,
  ``Graviton as a Pair of Collinear Gauge Bosons,''
Phys.\ Lett.\ B {\bf 739}, 457 (2014).
[arXiv:1409.4771 [hep-th]].}

\lref\notation{M.L.~Mangano and S.J.~Parke,
``Multiparton amplitudes in gauge theories,''
Phys. Rept.  {\bf 200}, 301 (1991).
[hep-th/0509223];\br
L.J.~Dixon,
  ``Calculating scattering amplitudes efficiently,''
in Boulder 1995, QCD and beyond 539-582.
[hep-ph/9601359].}

\lref\CachazoFWA{
  F.~Cachazo and A.~Strominger,
``Evidence for a New Soft Graviton Theorem,''
[arXiv:1404.4091 [hep-th]].
}

\lref\CasaliXPA{
  E.~Casali,
``Soft sub-leading divergences in Yang-Mills amplitudes,''
JHEP {\bf 1408}, 077 (2014).
[arXiv:1404.5551 [hep-th]].
}

\lref\toappear{
S. Stieberger and T.R. Taylor,
``Disk Scattering of Open and Closed Strings,''
MPP-2015--184, to appear.}

\Title{\vbox{\rightline{MPP--2015--183}
}}
{\vbox{\centerline{Subleading Terms in the Collinear Limit of}
\bigskip\centerline {Yang-Mills Amplitudes}
}}
\medskip
\centerline{Stephan Stieberger$^a$ and Tomasz R. Taylor$^b$}
\bigskip
\centerline{\it $^a$ Max--Planck--Institut f\"ur Physik}
\centerline{\it Werner--Heisenberg--Institut, 80805 M\"unchen, Germany}
\medskip
\centerline{\it  $^b$ Department of Physics}
\centerline{\it  Northeastern University, Boston, MA 02115, USA}

\vskip15pt

\medskip
\bigskip\bigskip\bigskip
\centerline{\bf Abstract}
\vskip .2in
\noindent

\noindent
{}For two massless particles $i$ and $\!j$, the collinear limit is a special kinematic configuration in which the particles propagate with parallel four-momentum vectors, with the total momentum $P$ distributed as $p_i=xP$ and $p_j=(1{-}x)P$, so that $s_{ij}\equiv (p_i+p_j)^2=P^2=0$. In Yang-Mills theory, if $i$ and $j$ are among $N$ gauge bosons participating in a scattering process, it is well known that the partial amplitudes associated to the (single trace) group factors with adjacent $i$ and $j$ are singular in the collinear limit and factorize at the leading order into $(N{-}1)$-particle amplitudes times the universal, $x$-dependent Altarelli-Parisi factors. We give a precise definition of the collinear limit and show that at the tree level, the subleading, non-singular terms are related to the amplitudes with a single graviton inserted instead of two collinear gauge bosons. To that end, we argue that in one-graviton Einstein-Yang-Mills amplitudes, the graviton with momentum $P$ can be replaced by a pair of collinear gauge bosons carrying arbitrary momentum fractions $xP$ and $(1{-}x)P$.

\Date{}
\noindent
\goodbreak
\break

\noindent
In high energy particle physics, collinear kinematics are very common. Viewed in the laboratory frame, all quarks and gluons (partons) propagating inside protons accelerated at the Large Hadron Collider (LHC) move in the beam direction, with a very little of transverse momentum. Since the early days of Quantum Chromodynamics (QCD), such collinear parton configurations have been in the focus of perturbative computations. In the so-called leading logarithmic approximation, the violation of Bjorken scaling \BjorkenDY\ in  proton structure functions can be understood as an effect of $1\to 2$ parton decays which are necessarily collinear. They are described by Altarelli-Parisi probabilities \AltarelliZS\ and involve the running gauge coupling constant that brings the fundamental QCD mass scale.  When protons collide at high energies, many quarks and gluons are often produced in a single two-parton collision. Multi-parton amplitudes
favor collinear final state configurations due to the singular behaviour that will be discussed below. At the LHC, such partons fragment into hadronic jets.

{}In general, for two massless particles $i$ and $\!j$, the collinear limit is defined as a special kinematic configuration in which the particles propagate with parallel four-momentum vectors, with the total momentum $P$ distributed as $p_i=xP$ and $p_j=(1-x)P$, so that $s_{ij}\equiv (p_i+p_j)^2=P^2=0$. In QCD, as in any Yang-Mills theory, if $i$ and $j$ are among $N$ gluons participating in a scattering process, it is well known that the partial amplitudes \notation\ associated to the (single trace) color factors with adjacent $i$ and $j$ are singular in the collinear limit and factorize at the leading order into $(N{-}1)$-gluon amplitudes times the universal, $x$-dependent Altarelli-Parisi factors (three-gluon MHV amplitudes). The singularity is a simple pole at $s_{ij}=0$ due to an intermediate gluon propagating on zero mass shell. In this paper, we go beyond the leading pole approximation and discuss non--factorizable, finite contributions\foot{The ``leading logarithms'' come from integrating such poles. In the language of perturbative QCD, the subleading terms discussed here belong to so-called ``higher twist'' contributions.}.

In order to give a precise definition of the leading and subleading parts, we need to specify how the collinear limit is reached from a generic kinematic configuration. Let us specify to generic light-like momenta $p_i=p_{N-1},~p_j=p_N$ and introduce two light-like vectors $P$ and $r$ such that the momentum spinors decompose as
\eqn\cols{\eqalign{\lambda_{N-1}&=\lambda_P\cos\theta-\epsilon\lambda_r\sin\theta\ ,
\quad\quad\quad\tilde\lambda_{N-1}=\tilde\lambda_P\cos\theta
-\epsilon\tilde\lambda_r\sin\theta\ ,
\cr
\lambda_{N}&=\lambda_P\sin\theta+\epsilon\lambda_r\cos\theta\ ,
\quad\quad\quad~~~\,\tilde\lambda_{N}=\tilde\lambda_P\sin\theta+
\epsilon\tilde\lambda_r\cos\theta\ ,
}}
hence
\eqn\pps{\eqalign{p_{N-1}& ={\bf c}^2P-\epsilon\,{\bf s c} (\lambda_P\tilde\lambda_r+
\lambda_r\tilde\lambda_P)+ \epsilon^2{\bf s}^2 r\ ,\cr
p_{N}& ={\bf s}^2P+\epsilon\,{\bf sc} (\lambda_P\tilde\lambda_r+
\lambda_r\tilde\lambda_P)+ \epsilon^2{\bf c}^2 r\ ,
}}
where
\eqn\nota{{\bf{c}}\equiv\cos\theta=\sqrt{x}~,\qquad\qquad {\bf{s}}\equiv\sin\theta=\sqrt{1-x}\ .}
We also have
\eqn\pron{\langle N{-}1\, N\rangle=\epsilon\,\langle Pr\rangle~, \qquad
[N{-}1\, N]=\epsilon\,[Pr]~.}
The total momentum is:
\eqn\colm{p_{N-1}+p_N=P+\epsilon^2r~,\qquad (p_{N-1}+p_N)^2\equiv s_{N{-}1, N}= 2 Pr\,\epsilon^2\ .}
The collinear configuration will be reached in the $\epsilon\to 0$ limit and
 the tree amplitudes discussed below will be expanded in powers of $\epsilon$. Partial amplitudes with adjacent $N{-}1$ and $N$ contain single (factorization) poles. Thus their leading terms
are of order ${\cal O}(\epsilon^{-1})$ and the subleading ones are  of order ${\cal O}(\epsilon^0)$. Collinear expansions of partial amplitudes with non-adjacent $N{-}1$ and $N$ start at the subleading ${\cal O}(\epsilon^0)$ order.

The leading collinear behaviour of amplitudes with adjacent $N{-}1,\, N$ is well known~\notation\ and depends on respective helicities. For identical helicities,
\eqn\ppsa{\eqalign{A(1,\dots,N{-}1^+, N^+)& ={1\over \langle N{-}1\, N\rangle\,{\bf sc}} A(1,\dots,P^+)+\epsilon^0 {\rm Sub}^{++}+\dots\ ,
\cr
A(1,\dots,N{-}1^-, N^-)& ={1\over [ N{-}1\, N]\,{\bf sc}} A(1,\dots,P^-)\,+\epsilon^0 {\rm Sub}^{--}+\dots\ ,
}}
where we used superscripts to denote helicity states. Here, Sub denote subleading contributions which are the focus of this work. The remaining terms vanish in the $\epsilon\to 0$ limit. For opposite helicities:
\eqn\ppsb{\eqalign{A(1,\dots,N{-}1^+, N^-)& ={{\bf s}^3\over \langle N{-}1\, N\rangle\,{\bf c}} A(1,\dots,P^-)
+{{\bf c}^3\over [ N{-}1\, N]\,{\bf s}} A(1,\dots,P^+) \cr &\qquad\qquad\qquad \qquad +\,\epsilon^0 {\rm Sub}^{+-}+\dots\ .
}}

The starting point for our discussion of subleading terms is the recent observation \StiebergerCEA\ that the tree-level Einstein-Yang-Mills amplitudes describing decays of a single graviton or a dilaton into a number of gauge bosons, can be written as linear combinations of pure gauge amplitudes in which the graviton (or dilaton) is replaced by a pair of gauge bosons. Their $\pm 1$ helicities add up to $\pm 2$ for the graviton or to 0 for the dilaton.  Each of them carries exactly one half of the original graviton or dilaton momentum, which is a special case of a collinear configuration with
${\bf s}={\bf c}=\sqrt{1/2}$. From now on we will focus on graviton amplitudes. The crucial point is that the relations derived in \StiebergerCEA\ can be extended to {\it arbitrary\/} collinear configurations, in the following way
\eqnn\FINALL{
$$\eqalignno{\quad A_{\rm EYM}&(1,2,\dots,N{-}2;P^{\pm 2})=&\FINALL\cr
\quad={\kappa\,{\bf s}^2\over g^2}&\Biggl\{ \!\sum_{l=2}^{\ceiling{\fc{N}{2}}-1}\sum_{i=2}^l
\Big(\sum_{j=i}^ls_{j,N-1}\Big)\ A_{\rm YM}(1,\ldots,i-1,N^{\pm},i,\dots,l,N{-}1^\pm,l+1,\ldots,N{-}2)\cr
\,\,+&\sum_{l=\ceiling{\fc{N}{2}}}^{N-3}\sum_{i=l+1}^{N-2}
\Big(\!\sum_{j=l+1}^is_{j,N-1}\Big)\ A_{\rm YM}(1,\ldots,l,N{-}1^\pm,l+1,\ldots,i,N^\pm,i+1,\ldots,N{-}2)\Biggr\} ,}
$$}
$\!\!$where $\kappa$ and $g$ are the gravitational and gauge coupling constants, respectively\foot{$\ceiling{{N\over 2}}$ is  the smallest integer greater than or equal to ${N\over 2}$. Since the graviton is identified by its momentum $P$, we can skip in the following the EYM and YM labelings of the amplitudes.}.  On the left hand side, we have a mixed gauge-gravitational amplitude involving a single graviton of momentum $P$, helicity $\pm 2$ as indicated by the superscript, and  $N{-}2$ gluons. This amplitude is associated to a single trace color factor with the respective gluon ordering.
On the right hand side, we have a linear combination of pure gauge, partial amplitudes weighted by the kinematic invariants $s_{j,N-1}=2p_jp_{N-1}$. Here, the graviton is replaced by two gluons in the collinear configuration:
\eqn\conf{p_{N-1}={\bf c}^2 P=xP~, \qquad p_{N}={\bf s}^2 P=(1-x)P\ ,}
 {\it i.e}.\ the leading ${\cal O}(\epsilon^0)$ order of Eqs. \cols\ and \pps.
 Note that on the right hand side, $N{-}1$ and $N$ are never adjacent, therefore the Einstein-Yang-Mills  amplitude emerges from the collinear limit  of Yang-Mills amplitudes
at the subleading ${\cal O}(\epsilon^0)$ order.
In order to further discuss Eq. \FINALL, it is useful to write it explicitly for $N=5,6,7$:
\eqnn\three
\eqnn\four
\eqnn\five
$$\eqalignno{A(1,2,3;P^{\pm 2})&={\kappa\,{\bf s}^2\over g^2}\ s_{24}\ A(1,5^\pm,2,4^\pm,3)\ ,&\three\cr
A(1,2,3,4;P^{\pm 2})&={\kappa\,{\bf s}^2\over g^2}\Big\{s_{25}\ A(1,6^\pm,2,5^\pm,3,4)+s_{45}\ A(1,2,3,5^\pm,4,6^\pm)\Big\}\ ,&\four\cr
A(1,2,3,4,5;P^{\pm 2})&={\kappa\,{\bf s}^2\over g^2}\Big\{s_{26}\ A(1,7^\pm,2,6^\pm,3,4,5)+s_{36}\  A(1,2,7^\pm,3,6^\pm,4,5)\cr
&+(s_{36}+s_{26})\ A(1,7^\pm,2,3,6^\pm,4,5)+s_{56}\ A(1,2,3,4,6^\pm,5,7^\pm)\Big\}\ .\ \ \ \ \ \ \ &\five}$$

The fact that the relations written in Ref. \StiebergerCEA\ can be extended from ${\bf s}={\bf c}=\sqrt{1/2}$, {\it i.e}.\ from $x=1/2$, to  an arbitrary collinear configuration by inserting a simple factor of ${\bf s}^2=1-x$ is highly non--trivial. It is easiest to check
for the helicity configurations  described by MHV amplitudes \ParkeGB\ on the r.h.s.\ of Eq. \FINALL, {\it i.e}.\ when the collinear pair is among $N-2$ gluons with identical helicities and there are only two gluons with opposite helicities\foot{The other case, when the collinear pair carry helicities opposite to all other $N{-}2$ gluons, does not contribute because the corresponding amplitudes vanish in the collinear limit as $\epsilon^4$.}. Then  the r.h.s.\ of Eq. \FINALL\ is a homogenous function of spinor (and momentum) variables and it is easy to see that, in this case, arbitrary value of $x$ can  be reached  from $x=1/2$ by a simple rescaling of the amplitudes, with the net effect of an overall $1-x$ factor. For other helicity configurations, the amplitudes are not homogenous in the momenta of collinear gluons. Already at the NMHV level, individual amplitudes contain poles in three-gluon channels $[ijN{-}1]$ and $ [ijN]$ (with $i,j\neq N{-}1,N$), characterized by the kinematic invariants
\eqn\tpol{\eqalign{ t_{ijN{-}1}   &\equiv(p_i+p_j+p_{N{-}1})^2={\bf c}^2 t_{ijP}+ {\bf s}^2 s_{ij}+\dots\cr
t_{ijN}   &\equiv(p_i+p_j+p_{N})^2={\bf s}^2 t_{ijP}+ {\bf c}^2 s_{ij}+\dots
}}
{}Such poles must cancel on the r.h.s.\ of Eq. \FINALL\ for the EYM amplitude to be free of unphysical singularities. In the appendix, we show that it is indeed the case for $N=6$, and obtain an explicit expression for $A(1^+,2^+,3^-,4^+;P^{-2})$ in agreement with Eq. \four. For $N=7$, a similar check is still possible but it involves very tedious computations. Starting from $N=8$, NNMHV amplitudes can appear on the r.h.s.\ of Eq. \FINALL, therefore a complete proof would have to rely on more general representation of tree amplitudes or on recursion relations. Actually, the most straightforward way is to consider these amplitudes as a zero--slope limit of superstring disk amplitudes involving open and closed strings. Then, Eqs. \FINALL\ can be proven for arbitary helicity configurations \toappear.
In this work however, we focus on field--theoretical amplitudes.

At the tree level, there are $(N-3)!$ independent $N$-gluon amplitudes \BernQJ. For a given $N$, we can express the Yang-Mills amplitudes appearing in Eq. \FINALL\ in terms of the  basis
$A(1,\sigma(2,3,\dots N{-}2),N{-}1,N)$, where $\sigma$ denotes the set of $(N-3)!$ permutations  of $2,3,\dots, N{-}2$. Let us start from $N=5$, as in Eq. \three, where we use:
\eqn\bcj{A(1,5,2,4,3)={s_{21}\over s_{25}}\ A(1,2,3,4,5)+
{s_{21}+s_{23}\over s_{25}}\  A(1,3,2,4,5)\ .}
As a result, we conclude that the following relation is valid up to the ${\cal O}(\epsilon^0)$ order:
\eqn\bbcj{s_{3P}A(1,2,3,4^\pm,5^\pm)-s_{2P}A(1,3,2,4^\pm,5^\pm)=
{g^2\over \kappa\,x}A(1,2,3;P^{\pm 2})\ .}
By using  Eq. \ppsa\ and BCJ relations for four-gluon amplitudes \BernQJ, it is easy to see that the leading collinear singularities ${\cal O}(\epsilon^{-1})$ drop out, therefore Eq. \bbcj\ connects the subleading terms with the mixed gauge-gravitational amplitude. For $N=5$, we obtain one relation between the subleading parts of two independent amplitudes. For $N=6$, a similar equation reads:
\eqn\abcj{\eqalign{s_{4P}A(1,2,3,4,&\ 5^\pm,6^\pm)\,-\,s_{3P}[A(1,2,4,3,5^\pm,6^\pm)+A(1,4,2,3,5^\pm,6^\pm)]
\cr
&+s_{2P}A(1,4,3,2,5^\pm,6^\pm)
={g^2\over \kappa\,x}A(1,2,3,4;P^{\pm 2})\ .}}
In this case, however, we have two additional mixed amplitudes,  say $A(1,3,2,4;P^{\pm 2})$ and $A(1,2,4,3;P^{\pm 2})$, that can be used in similar relations, obtained by interchanging $2\leftrightarrow 3$ and $3\leftrightarrow 4$, respectively. As a result, we obtain three relations for the subleading parts of six independent gauge amplitudes\foot{Three other mixed amplitudes are related by parity reflections, therefore they do not provide additional constraints.}. For $N=7$,
\eqn\abcj{\eqalign{s_{5P}&A(1,2,3,4,5, 6^\pm,7^\pm)\cr -&\ s_{4P}[A(1,2,3,5,4,6^\pm,7^\pm)+A(1,2,5,3,4,6^\pm,7^\pm)+A(1,5,2,3,4,6^\pm,7^\pm)]\cr +&\ s_{3P}[A(1,5,4,2,3,6^\pm,7^\pm)+A(1,5,2,4,3,6^\pm,7^\pm)+A(1,2,5,4,3,6^\pm,7^\pm)]\cr
 -&\ s_{2P}A(1,5,4,3,2,6^\pm,7^\pm)
~=~ {g^2\over \kappa\,x}A(1,2,3,4,5;P^{\pm 2})\ .}}
In this case, there are 24 independent Yang-Mills amplitudes with the subleading collinear behaviour constrained by twelve Einstein-Yang-Mills amplitudes. 
For arbitrary $N$ a similar formula reads
\eqn\similar{
\sum_{\rho\in\Pc_N} (-1)^{m_\rho}\ s_{\rho(N-2)P}\ A(1,\rho(2,\ldots,N-2),N-1,N)
={g^2\over \kappa\,x}A(1,\ldots,N-2;P)\ ,}
where $\Pc_N$ is a subset of permutations acting on $2,\ldots,N-2$
and $m_{\rho}\in\{0,1\}$ as specified in \toappear. Now there are $(N-3)!/2$ independent constraints.

We see that the subleading collinear behaviour of pure gauge amplitudes is  determined in part by the amplitudes with the graviton inserted instead of the collinear pair. Twice as many constraints are necessary, however, in order to fully determine the subleading terms for all amplitudes. In another physically interesting case of soft ($x\to 0$) divergences, the subleading behaviour has been recently discussed in Einstein's gravity \CachazoFWA\ and in Yang-Mills theory \CasaliXPA. We hope that similar considerations will allow complete determination of the subleading behaviour in the collinear case.

The fact that the graviton can be replaced by two gluons in arbitrary collinear configurations in the single-graviton amplitudes of Eq. \FINALL\ raises an interesting question whether pure Einstein, multi-graviton amplitudes share this property. The recent linearization \StiebergerQJA\ of KLT relations \KawaiXQ\ suggests that this may be the case. It would be another indication for the existence of some underlying gauge structure in quantum gravity.

\vskip0.5cm
\goodbreak
\leftline{\noindent{\it Appendix}}

\noindent
We will show that Eq. \four\ holds for $A(1^+,2^+,3^-,4^+;P^{-2})$. To that end, we take the collinear limits, c.f.\ Eq. \cols, of the six-gluon NMHV amplitudes written in Ref. \notation:
\eqn\ammone{\eqalign{A(1^+,6^-,2^+,5^-,3^-,4^+)={\langle 3P\rangle^4\over s_{1P}s_{2P}s_{3P}}\Bigg\{&{[14]^2[23]^2\over {\bf c}^2({\bf s}^2s_{23}+{\bf c}^2s_{14})s_{14}} +{[12]^2[34]^2\over{\bf s}^2 ({\bf s}^2s_{34}+{\bf c}^2s_{12})s_{34}}\cr &\qquad+\, {[12][23][34][41]\over{\bf c}^2{\bf s}^2 s_{14}s_{34}}\Bigg\}+\dots }}
\eqn\ammtwo{\eqalign{A(1^+,2^+,3^-,5^-,4^+,6^-)={\langle 3P\rangle^4\over s_{1P}s_{3P}s_{4P}}\Bigg\{&{[14]^2[23]^2\over {\bf s}^2({\bf s}^2s_{23}+{\bf c}^2s_{14})s_{23}} +{[12]^2[34]^2\over{\bf c}^2 ({\bf s}^2s_{34}+{\bf c}^2s_{12})s_{12}}\cr &\qquad+\, {[12][23][34][41]\over{\bf c}^2{\bf s}^2 s_{12}s_{23}}\Bigg\}+\dots }}
where we omitted terms that vanish in the $\epsilon\to 0$ limit; we also set $g^2=\kappa=1$.
After substituting into Eq. \four\ and using momentum conservation, we obtain
\eqn\eymf{A(1^+,2^+,3^-,4^+;P^{-2})={\langle 3P\rangle^4\over \langle 12\rangle\langle 23\rangle\langle 34\rangle\langle 41\rangle}\ ,}
in agreement with Ref. \StiebergerCEA.
For other NMHV helicity configurations, Eq. \four\ follows in exactly the same way.

\vskip0.5cm
\goodbreak
\leftline{\noindent{\bf Acknowledgments}}

\noindent
TRT is grateful to CERN Theory Unit and to Max-Planck-Institut f\"ur Physik in M\"unchen,  where substantial portions this work were performed, for their hospitality and financial support.
This material is based in part upon work supported by the National Science Foundation under Grant No.\ PHY-1314774.   Any
opinions, findings, and conclusions or recommendations expressed in
this material are those of the authors and do not necessarily reflect
the views of the National Science Foundation.

\listrefs

\end
{\it Since $p_1=q_1$ (with $\mu_1=\nu_1$) for adjacent gluons, in order to obtain a well-defined amplitude, we start from $p_1\neq q_1$ and take the limit after symmetrizing in $\{p_1,q_1\}$, thus removing the leading collinear singularity. By using BCJ relations \BernQJ, such a symmetric combination can always be rewritten in terms of manifestly finite amplitudes in which the collinear gluons appear at non--adjacent positions, {\it cf}.\ in the example discussed at the end of the paper.} {\bf Is this paragraph still an issue - only for $N=3$, so perhaps we can move this comment to the example ?}
\eqn\adef{\eqalign{A[&p,N{-}1,N{-}2,\dots,2, 1,1,2,\dots,N{-}2,N{-}1,q]\cr &\equiv
A[p,\mu={+}1;p_{N{-}1},\mu_{N{-}1};\dots;p_1,\mu_1; q_1,\nu_1;\dots;q_{N{-}1},\nu_{N{-}1};
q,\nu={-}1]~.
}}

%% file: myharvmac
%
%
%
\def\unredoffs{} \def\redoffs{\voffset=-.31truein\hoffset=-.48truein}
\def\speclscape{}
%
%
%
%
%
\newbox\leftpage \newdimen\fullhsize \newdimen\hstitle \newdimen\hsbody
\tolerance=1000\hfuzz=2pt
\catcode`\@=11 
\ifx\hyperdef\UNd@FiNeD\def\hyperdef#1#2#3#4{#4}\def\hyperref#1#2#3#4{#4}\fi
\def\bigans{b }
\def\answ{b }
%
\ifx\answ\bigans\message{(This will come out unreduced.}
\magnification=1200\unredoffs\baselineskip=16pt plus 2pt minus 1pt
\hsbody=\hsize \hstitle=\hsize 
\else\message{(This will be reduced.} \let\l@r=L
\magnification=1000\baselineskip=16pt plus 2pt minus 1pt \vsize=7truein
\redoffs \hstitle=8truein\hsbody=4.75truein\fullhsize=10truein\hsize=\hsbody
\output={\ifnum\pageno=0 
  \shipout\vbox{\speclscape{\hsize\fullhsize\makeheadline}
    \hbox to \fullhsize{\hfill\pagebody\hfill}}\advancepageno
  \else
  \almostshipout{\leftline{\vbox{\pagebody\makefootline}}}\advancepageno
  \fi}
\def\almostshipout#1{\if L\l@r \count1=1 \message{[\the\count0.\the\count1]}
      \global\setbox\leftpage=#1 \global\let\l@r=R
 \else \count1=2
  \shipout\vbox{\speclscape{\hsize\fullhsize\makeheadline}
      \hbox to\fullhsize{\box\leftpage\hfil#1}}  \global\let\l@r=L\fi}
\fi
%
\newcount\yearltd\yearltd=\year\advance\yearltd by -2000

\def\Title#1#2{\nopagenumbers\abstractfont\hsize=\hstitle\rightline{#1}%
\vskip 1in\centerline{\titlefont #2}\abstractfont\vskip .5in\pageno=0}
\def\Date#1{\vfill\leftline{#1}\tenpoint\supereject\global\hsize=\hsbody%
\footline={\hss\tenrm\hyperdef\hypernoname{page}\folio\folio\hss}}%
%

\def\draftmode{\message{ DRAFTMODE }\def\draftdate{{\rm preliminary draft:
\number\month/\number\day/\number\yearltd\ \ \hourmin}}%
\headline={\hfil\draftdate}\writelabels\baselineskip=20pt plus 2pt minus 2pt
 {\count255=\time\divide\count255 by 60 \xdef\hourmin{\number\count255}
  \multiply\count255 by-60\advance\count255 by\time
  \xdef\hourmin{\hourmin:\ifnum\count255<10 0\fi\the\count255}}}
\def\nolabels{\def\wrlabeL##1{}\def\eqlabeL##1{}\def\reflabeL##1{}}
\def\writelabels{\def\wrlabeL##1{\leavevmode\vadjust{\rlap{\smash%
{\line{{\escapechar=` \hfill\rlap{\sevenrm\hskip.03in\string##1}}}}}}}%
\def\eqlabeL##1{{\escapechar-1\rlap{\sevenrm\hskip.05in\string##1}}}%
\def\reflabeL##1{\noexpand\llap{\noexpand\sevenrm\string\string\string##1}}}
\nolabels
%
\global\newcount\secno \global\secno=0
\global\newcount\meqno \global\meqno=1
\def\s@csym{}
\def\newsec#1{\global\advance\secno by1%
{\toks0{#1}\message{(\the\secno. \the\toks0)}}%
\global\subsecno=0\eqnres@t\let\s@csym\secsym\xdef\secn@m{\the\secno}\noindent
{\bf\hyperdef\hypernoname{section}{\the\secno}{\the\secno.} #1}%
\writetoca{{\string\hyperref{}{section}{\the\secno}{\it\the\secno.}} {{\it #1} }}%
\par\nobreak\medskip\nobreak}
\def\eqnres@t{\xdef\secsym{\the\secno.}\global\meqno=1\bigbreak\bigskip}
\def\sequentialequations{\def\eqnres@t{\bigbreak}}\xdef\secsym{}
\global\newcount\subsecno \global\subsecno=0
\def\subsec#1{\global\advance\subsecno by1%
{\toks0{#1}\message{(\s@csym\the\subsecno. \the\toks0)}}%
\ifnum\lastpenalty>9000\else\bigbreak\fi       \global\subsubsecno=0
\noindent{\it\hyperdef\hypernoname{subsection}{\secn@m.\the\subsecno}%
{\secn@m.\the\subsecno.} #1}\writetoca{\string\quad
{\string\hyperref{}{subsection}{\secn@m.\the\subsecno}{\secn@m.\the\subsecno.}}
{#1}}\par\nobreak\medskip\nobreak}
\def\appendix#1#2{\global\meqno=1\global\subsecno=0\xdef\secsym{\hbox{#1.}}%
\bigbreak\bigskip\noindent{\bf Appendix \hyperdef\hypernoname{appendix}{#1}%
{#1.} #2}{\toks0{(#1. #2)}\message{\the\toks0}}%
\xdef\s@csym{#1.}\xdef\secn@m{#1}%
\writetoca{\string\hyperref{}{appendix}{#1}{{\it Appendix} {\it #1.}} {\it #2}}%
\par\nobreak\medskip\nobreak}
%
%
\def\checkm@de#1#2{\ifmmode{\def\f@rst##1{##1}\hyperdef\hypernoname{equation}%
{#1}{#2}}\else\hyperref{}{equation}{#1}{#2}\fi}
\def\eqnn#1{\DefWarn#1\xdef #1{(\noexpand\relax\noexpand\checkm@de%
{\s@csym\the\meqno}{\secsym\the\meqno})}%
\wrlabeL#1\writedef{#1\leftbracket#1}\global\advance\meqno by1}
\def\f@rst#1{\c@t#1a\em@ark}\def\c@t#1#2\em@ark{#1}
\def\eqna#1{\DefWarn#1\wrlabeL{#1$\{\}$}%
\xdef #1##1{(\noexpand\relax\noexpand\checkm@de%
{\s@csym\the\meqno\noexpand\f@rst{##1}}{\hbox{$\secsym\the\meqno##1$}})}
\writedef{#1\numbersign1\leftbracket#1{\numbersign1}}\global\advance\meqno by1}
\def\eqn#1#2{\DefWarn#1%
\xdef #1{(\noexpand\hyperref{}{equation}{\s@csym\the\meqno}%
{\secsym\the\meqno})}$$#2\eqno(\hyperdef\hypernoname{equation}%
{\s@csym\the\meqno}{\secsym\the\meqno})\eqlabeL#1$$%
\writedef{#1\leftbracket#1}\global\advance\meqno by1}
\def\xeqn{\expandafter\xe@n}\def\xe@n(#1){#1}
\def\xeqna#1{\expandafter\xe@n#1}
\def\eqns#1{(\e@ns #1{\hbox{}})}
\def\e@ns#1{\ifx\UNd@FiNeD#1\message{eqnlabel \string#1 is undefined.}%
\xdef#1{(?.?)}\fi{\let\hyperref=\relax\xdef\next{#1}}%
\ifx\next\em@rk\def\next{}\else%
\ifx\next#1\xeqn#1\else\def\n@xt{#1}\ifx\n@xt\next#1\else\xeqna#1\fi
\fi\let\next=\e@ns\fi\next}

\def\DefWarn#1{\ifx\UNd@FiNeD#1\else
\immediate\write16{*** WARNING: the label \string#1 is already defined ***}\fi}
%
\newskip\footskip\footskip14pt plus 1pt minus 1pt 
\def\footnotefont{\ninepoint}\def\f@t#1{\footnotefont #1\@foot}
\def\f@@t{\baselineskip\footskip\bgroup\footnotefont\aftergroup\@foot\let\next}
\setbox\strutbox=\hbox{\vrule height9.5pt depth4.5pt width0pt}
\global\newcount\ftno \global\ftno=0
\def\foot{\global\advance\ftno by1\def\foot@rg{\hyperref{}{footnote}%
{\the\ftno}{\the\ftno}\xdef\foot@rg{\noexpand\hyperdef\noexpand\hypernoname%
{footnote}{\the\ftno}{\the\ftno}}}\footnote{$^{\foot@rg}$}}
%
\newwrite\ftfile
\def\footend{\def\foot{\global\advance\ftno by1\chardef\wfile=\ftfile
\hyperref{}{footnote}{\the\ftno}{$^{\the\ftno}$}%
\ifnum\ftno=1\immediate\openout\ftfile=\jobname.fts\fi%
\immediate\write\ftfile{\noexpand\smallskip%
\noexpand\item{\noexpand\hyperdef\noexpand\hypernoname{footnote}
{\the\ftno}{f\the\ftno}:\ }\pctsign}\findarg}%
\def\footatend{\vfill\eject\immediate\closeout\ftfile{\parindent=20pt
\centerline{\bf Footnotes}\nobreak\bigskip\input \jobname.fts }}}
\def\footatend{}
%
%
\global\newcount\refno \global\refno=1
\newwrite\rfile
\def\ref{[\hyperref{}{reference}{\the\refno}{\the\refno}]\nref}
\def\nref#1{\DefWarn#1%
\xdef#1{[\noexpand\hyperref{}{reference}{\the\refno}{\the\refno}]}%
\writedef{#1\leftbracket#1}%
\ifnum\refno=1\immediate\openout\rfile=\jobname.refs\fi
\chardef\wfile=\rfile\immediate\write\rfile{\noexpand\item{[\noexpand\hyperdef%
\noexpand\hypernoname{reference}{\the\refno}{\the\refno}]\ }%
\reflabeL{#1\hskip.31in}\pctsign}\global\advance\refno by1\findarg}
\def\findarg#1#{\begingroup\obeylines\newlinechar=`\^^M\pass@rg}
{\obeylines\gdef\pass@rg#1{\writ@line\relax #1^^M\hbox{}^^M}%
\gdef\writ@line#1^^M{\expandafter\toks0\expandafter{\striprel@x #1}%
\edef\next{\the\toks0}\ifx\next\em@rk\let\next=\endgroup\else\ifx\next\empty%
\else\immediate\write\wfile{\the\toks0}\fi\let\next=\writ@line\fi\next\relax}}
\def\striprel@x#1{} \def\em@rk{\hbox{}}
\def\lref{\begingroup\obeylines\lr@f}
\def\lr@f#1#2{\DefWarn#1\gdef#1{\let#1=\UNd@FiNeD\ref#1{#2}}\endgroup\unskip}

\def\addref#1{\immediate\write\rfile{\noexpand\item{}#1}} 
\def\listrefs{\footatend\vfill\supereject\immediate\closeout\rfile\writestoppt
\baselineskip=\footskip\centerline{{\bf References}}\bigskip{\parindent=20pt%
\frenchspacing\escapechar=` \input \jobname.refs\vfill\eject}\nonfrenchspacing}
\def\startrefs#1{\immediate\openout\rfile=\jobname.refs\refno=#1}
\def\xref{\expandafter\xr@f}\def\xr@f[#1]{#1}
\def\refs#1{\count255=1[\r@fs #1{\hbox{}}]}
\def\r@fs#1{\ifx\UNd@FiNeD#1\message{reflabel \string#1 is undefined.}%
\nref#1{need to supply reference \string#1.}\fi%
\vphantom{\hphantom{#1}}{\let\hyperref=\relax\xdef\next{#1}}%
\ifx\next\em@rk\def\next{}%
\else\ifx\next#1\ifodd\count255\relax\xref#1\count255=0\fi%
\else#1\count255=1\fi\let\next=\r@fs\fi\next}
%

%
\newwrite\ffile\global\newcount\figno \global\figno=1
\def\fig{fig.~\hyperref{}{figure}{\the\figno}{\the\figno}\nfig}
\def\nfig#1{\DefWarn#1%
\xdef#1{fig.~\noexpand\hyperref{}{figure}{\the\figno}{\the\figno}}%
\writedef{#1\leftbracket fig.\noexpand~\xfig#1}%
\ifnum\figno=1\immediate\openout\ffile=\jobname.figs\fi\chardef\wfile=\ffile%
{\let\hyperref=\relax
\immediate\write\ffile{\noexpand\medskip\noexpand\item{Fig.\ %
\noexpand\hyperdef\noexpand\hypernoname{figure}{\the\figno}{\the\figno}. }
\reflabeL{#1\hskip.55in}\pctsign}}\global\advance\figno by1\findarg}
\def\listfigs{\vfill\eject\immediate\closeout\ffile{\parindent40pt
\baselineskip14pt\centerline{{\bf Figure Captions}}\nobreak\medskip
\escapechar=` \input \jobname.figs\vfill\eject}}
\def\xfig{\expandafter\xf@g}\def\xf@g fig.\penalty\@M\ {}
\def\figs#1{figs.~\f@gs #1{\hbox{}}}
\def\f@gs#1{{\let\hyperref=\relax\xdef\next{#1}}\ifx\next\em@rk\def\next{}\else
\ifx\next#1\xfig #1\else#1\fi\let\next=\f@gs\fi\next}
\def\figin{\epsfcheck\figin}\def\figins{\epsfcheck\figins}
\def\epsfcheck{\ifx\epsfbox\UNd@FiNeD
\message{(NO epsf.tex, FIGURES WILL BE IGNORED)}
\gdef\figin##1{\vskip2in}\gdef\figins##1{\hskip.5in}
\else\message{(FIGURES WILL BE INCLUDED)}%
\gdef\figin##1{##1}\gdef\figins##1{##1}\fi}
\def\DefWarn#1{}
\def\figinsert{\goodbreak\midinsert}
\def\ifig#1#2#3{\DefWarn#1\xdef#1{Fig.~\noexpand\hyperref{}{figure}%
{\the\figno}{\the\figno}}\writedef{#1\leftbracket fig.\noexpand~\xfig#1}%
\figinsert\figin{\centerline{#3}}\medskip\centerline{\vbox{\baselineskip12pt
\advance\hsize by -1truein\noindent\wrlabeL{#1=#1}\footnotefont%
{\bf Fig.~\hyperdef\hypernoname{figure}{\the\figno}{\the\figno}:} #2}}
\bigskip\endinsert\global\advance\figno by1}
\newwrite\lfile
{\escapechar-1\xdef\pctsign{\string\%}\xdef\leftbracket{\string\{}
\xdef\rightbracket{\string\}}\xdef\numbersign{\string\#}}
\def\writedefs{\immediate\openout\lfile=\jobname.defs \def\writedef##1{%
{\let\hyperref=\relax\let\hyperdef=\relax\let\hypernoname=\relax
 \immediate\write\lfile{\string\def\string##1\rightbracket}}}}%
\def\writestop{\def\writestoppt{\immediate\write\lfile{\string\pageno
 \the\pageno\string\startrefs\leftbracket\the\refno\rightbracket
 \string\def\string\secsym\leftbracket\secsym\rightbracket
 \string\secno\the\secno\string\meqno\the\meqno}\immediate\closeout\lfile}}
\def\writestoppt{}\def\writedef#1{}
\def\seclab#1{\DefWarn#1%
\xdef #1{\noexpand\hyperref{}{section}{\the\secno}{\the\secno}}%
\writedef{#1\leftbracket#1}\wrlabeL{#1=#1}}
\def\subseclab#1{\DefWarn#1%
\xdef #1{\noexpand\hyperref{}{subsection}{\secn@m.\the\subsecno}%
{\secn@m.\the\subsecno}}\writedef{#1\leftbracket#1}\wrlabeL{#1=#1}}
\def\applab#1{\DefWarn#1%
\xdef #1{\noexpand\hyperref{}{appendix}{\secn@m}{\secn@m}}%
\writedef{#1\leftbracket#1}\wrlabeL{#1=#1}}
\newwrite\tfile \def\writetoca#1{}
\def\leaderfill{\leaders\hbox to 1em{\hss.\hss}\hfill}
\def\writetoc{\immediate\openout\tfile=\jobname.toc
   \def\writetoca##1{{\edef\next{\write\tfile{\noindent ##1
   \string\leaderfill {\string\hyperref{}{page}{\noexpand\number\pageno}%
                       {\noexpand\number\pageno}} \par}}\next}}}
\newread\ch@ckfile
\def\listtoc{\immediate\closeout\tfile\immediate\openin\ch@ckfile=\jobname.toc
\ifeof\ch@ckfile\message{no file \jobname.toc, no table of contents this pass}%
\else\closein\ch@ckfile\centerline{\bf Contents}\nobreak\medskip%
{\baselineskip=18.5pt  \footnotefont
\parskip=2pt\catcode`\@=12\input\jobname.toc
\catcode`\@=12\bigbreak\bigskip}\fi}
\catcode`\@=12 
%
\edef\tfontsize{\ifx\answ\bigans scaled\magstep3\else scaled\magstep4\fi}
\font\titlerm=cmr10 \tfontsize \font\titlerms=cmr7 \tfontsize
\font\titlermss=cmr5 \tfontsize \font\titlei=cmmi10 \tfontsize
\font\titleis=cmmi7 \tfontsize \font\titleiss=cmmi5 \tfontsize
\font\titlesy=cmsy10 \tfontsize \font\titlesys=cmsy7 \tfontsize
\font\titlesyss=cmsy5 \tfontsize \font\titleit=cmti10 \tfontsize
\skewchar\titlei='177 \skewchar\titleis='177 \skewchar\titleiss='177
\skewchar\titlesy='60 \skewchar\titlesys='60 \skewchar\titlesyss='60
\def\titlefont{\def\rm{\fam0\titlerm}
\textfont0=\titlerm \scriptfont0=\titlerms \scriptscriptfont0=\titlermss
\textfont1=\titlei \scriptfont1=\titleis \scriptscriptfont1=\titleiss
\textfont2=\titlesy \scriptfont2=\titlesys \scriptscriptfont2=\titlesyss
\textfont\itfam=\titleit \def\it{\fam\itfam\titleit}\rm}
 \ifx\answ\bigans\else scaled\magstep1\fi
\ifx\answ\bigans\def\abstractfont{\tenpoint}\else
\font\absit=cmti10 scaled \magstep1
\font\abssl=cmsl10 scaled \magstep1
\font\absrm=cmr10 scaled\magstep1 \font\absrms=cmr7 scaled\magstep1
\font\absrmss=cmr5 scaled\magstep1 \font\absi=cmmi10 scaled\magstep1
\font\absis=cmmi7 scaled\magstep1 \font\absiss=cmmi5 scaled\magstep1
\font\abssy=cmsy10 scaled\magstep1 \font\abssys=cmsy7 scaled\magstep1
\font\abssyss=cmsy5 scaled\magstep1 \font\absbf=cmbx10 scaled\magstep1
\skewchar\absi='177 \skewchar\absis='177 \skewchar\absiss='177
\skewchar\abssy='60 \skewchar\abssys='60 \skewchar\abssyss='60
\def\abstractfont{\def\rm{\fam0\absrm}
\textfont0=\absrm \scriptfont0=\absrms \scriptscriptfont0=\absrmss
\textfont1=\absi \scriptfont1=\absis \scriptscriptfont1=\absiss
\textfont2=\abssy \scriptfont2=\abssys \scriptscriptfont2=\abssyss
\textfont\itfam=\absit \def\it{\fam\itfam\absit}\def\footnotefont{\tenpoint}%
\textfont\slfam=\abssl \def\sl{\fam\slfam\abssl}%
\textfont\bffam=\absbf \def\bf{\fam\bffam\absbf}\rm}\fi
\def\tenpoint{\def\rm{\fam0\tenrm}
\textfont0=\tenrm \scriptfont0=\sevenrm \scriptscriptfont0=\fiverm
\textfont1=\teni  \scriptfont1=\seveni  \scriptscriptfont1=\fivei
\textfont2=\tensy \scriptfont2=\sevensy \scriptscriptfont2=\fivesy
\textfont\itfam=\tenit \def\it{\fam\itfam\tenit}\def\footnotefont{\ninepoint}%
\textfont\bffam=\tenbf \def\bf{\fam\bffam\tenbf}\def\sl{\fam\slfam\tensl}\rm}
\font\ninerm=cmr9 \font\sixrm=cmr6 \font\ninei=cmmi9 \font\sixi=cmmi6
\font\ninesy=cmsy9 \font\sixsy=cmsy6 \font\ninebf=cmbx9
\font\nineit=cmti9 \font\ninesl=cmsl9 \skewchar\ninei='177
\skewchar\sixi='177 \skewchar\ninesy='60 \skewchar\sixsy='60
\def\ninepoint{\def\rm{\fam0\ninerm}
\textfont0=\ninerm \scriptfont0=\sixrm \scriptscriptfont0=\fiverm
\textfont1=\ninei \scriptfont1=\sixi \scriptscriptfont1=\fivei
\textfont2=\ninesy \scriptfont2=\sixsy \scriptscriptfont2=\fivesy
\textfont\itfam=\ninei \def\it{\fam\itfam\nineit}\def\sl{\fam\slfam\ninesl}%
\textfont\bffam=\ninebf \def\bf{\fam\bffam\ninebf}\rm}
%
%
\def\noblackbox{\overfullrule=0pt}
\hyphenation{anom-aly anom-alies coun-ter-term coun-ter-terms}
\def\inv{^{\raise.15ex\hbox{${\scriptscriptstyle -}$}\kern-.05em 1}}

\def\Dsl{\,\raise.15ex\hbox{/}\mkern-13.5mu D} 
\def\dsl{\raise.15ex\hbox{/}\kern-.57em\partial}

\def\lspace{\ifx\answ\bigans{}\else\qquad\fi}
\def\lbspace{\ifx\answ\bigans{}\else\hskip-.2in\fi} 
\def\boxeqn#1{\vcenter{\vbox{\hrule\hbox{\vrule\kern3pt\vbox{\kern3pt
	\hbox{${\displaystyle #1}$}\kern3pt}\kern3pt\vrule}\hrule}}}
\def\mbox#1#2{\vcenter{\hrule \hbox{\vrule height#2in
		\kern#1in \vrule} \hrule}}  
%

\def\darr#1{\raise1.5ex\hbox{$\leftrightarrow$}\mkern-16.5mu #1}

\def\roughly#1{\raise.3ex\hbox{$#1$\kern-.75em\lower1ex\hbox{$\sim$}}}

\global\newcount\subsubsecno \global\subsubsecno=0
\def\subsubsec#1{\global\advance\subsubsecno by1%
{\toks0{#1}\message{(\the\secno\the\subsecno\the\subsubsecno. \the\toks0)}}%
\ifnum\lastpenalty>9000\else\bigbreak\fi
\noindent{\it\hyperdef\hypernoname{subsubsection}{\the\secno.\the\subsecno\the\subsubsecno}%
{\the\secno.\the\subsecno.\the\subsubsecno.} #1}
\par\nobreak\medskip\nobreak}
\def\boxit#1{\vbox{\hrule\hbox{\vrule\kern8pt
\vbox{\hbox{\kern8pt}\hbox{\vbox{#1}}\hbox{\kern8pt}}
\kern8pt\vrule}\hrule}}
\def\mathboxit#1{\vbox{\hrule\hbox{\vrule\kern8pt\vbox{\kern8pt
\hbox{$\displaystyle #1$}\kern8pt}\kern8pt\vrule}\hrule}}
\def\slashchar#1{\setbox0=\hbox{$#1$}           
   \dimen0=\wd0                                 
   \setbox1=\hbox{/} \dimen1=\wd1               
   \ifdim\dimen0>\dimen1                        
      \rlap{\hbox to \dimen0{\hfil/\hfil}}      
      #1                                        
   \else                                        
      \rlap{\hbox to \dimen1{\hfil$#1$\hfil}}   
      /                                         
   \fi}
\def\sqr#1#2{{\vcenter{\vbox{\hrule height.#2pt
         \hbox{\vrule width.#2pt height#1pt \kern#1pt
            \vrule width.#2pt}
         \hrule height.#2pt}}}}
